\documentclass[aps,twocolumn,showpacs,floatfix]{revtex4}
\usepackage{psfig}

\newcommand{\lsim}{\lower.7ex\hbox{$
    \;\stackrel{\textstyle<}{\sim}\;$}}
\def\order#1{{\cal O}\left(#1\right)}
\def\ep{\epsilon}
\newcommand{\be}{\begin{equation}}
\newcommand{\ee}{\end{equation}}
\newcommand{\ba}{\begin{eqnarray}}
\newcommand{\ea}{\end{eqnarray}}

\begin{document}

\title{
%
%
\[ \vspace{-2cm} \]
\noindent\hfill\hbox{\rm  } \vskip 1pt
\noindent\hfill\hbox{\rm Alberta Thy 17-01} \vskip 1pt
\noindent\hfill\hbox{\rm SLAC-PUB-9101} \vskip 1pt
\noindent\hfill\hbox{\rm hep-ph/0112264} \vskip 10pt
%
%
Semileptonic $b\to u$ decays: lepton invariant mass spectrum
}

\author{Andrzej Czarnecki}
\affiliation{
Department of Physics, University of Alberta\\
Edmonton, AB\ \  T6G 2J1, Canada\\
E-mail:  czar@phys.ualberta.ca}

\author{Kirill Melnikov}
\affiliation{Stanford Linear Accelerator Center\\
Stanford University, Stanford, CA 94309\\
E-mail: melnikov@slac.stanford.edu}

\begin{abstract}
We compute ${\cal O}(\alpha_s^2)$ QCD corrections to the lepton
invariant mass spectrum in the decay $b \to u l \nu_l$, relevant for
the determination of the CKM matrix element $|V_{ub}|$.  Our method
can also be used to evaluate moments of the lepton energy distribution
with an ${\cal O}(\alpha_s^2)$ accuracy.
The abelian part of our result gives the neutrino invariant mass
spectrum in the muon decay and, upon integration, the 
$\order{\alpha^2}$ correction to the muon lifetime.
\end{abstract}

\maketitle

Determination of the CKM matrix elements from precision studies in
$B$-physics is one of the main goals of experiments BaBar and
Belle, under way at SLAC and KEK.  These studies are expected to
provide important insights into flavor physics, in particular to shed
light on the origin of the CP-violation and, possibly, discover ``New
Physics.''

The two CKM parameters directly accessible at $B$-factories,
$|V_{cb}|$ and $|V_{ub}|$, strongly differ in magnitude, with the
former being about ten times larger than the latter.  An accurate
determination of $|V_{cb}|$ is much easier since the relevant decay
rates are relatively large and the backgrounds are small.  For
$|V_{ub}|$, the theoretically favorable methods are not feasible
experimentally, whereas interpretation of clean experimental
signatures suffers from large theoretical uncertainties.

Extraction of $|V_{ub}|$ from inclusive semileptonic decays of $B$
mesons requires a suppression of the much larger contribution of $b
\to c$ transitions. In order to do so one has to impose cuts on
various observables and several options have been discussed in the
literature. For example, one can select events with large energy of
the charged lepton, which can be produced only in $b\to u$ decays, or
require that the hadron invariant mass be smaller than the lightest
charmed meson $D$.

Unfortunately, such cuts are so severe that the rate of the remaining
events cannot be predicted using the Heavy Quark Expansion.  For
example, imposing the cut on the electron energy induces a sensitivity
of the decay rate to the $B$-meson light-cone wave function which is
not very well known.  It can in principle be extracted from
measurements of the photon energy spectrum in $b \to s \gamma$.
However, the relevant theoretical analysis has only been performed in
the limit of an infinite $b$ quark mass and the potentially sizable
$\Lambda_{\rm QCD}/m_b$ corrections are not under control.  It is
desirable, therefore, to have an alternative combination of cuts which
can remove the charm background, keep a significant fraction of $b \to
u$ events, and preserve the applicability of the standard Heavy Quark
Expansion.

Recently, a method fulfilling these requirements has been proposed by
Bauer, Ligeti and Luke \cite{Bauer:2000xf,Bauer:2001rc}. Their idea
consists in extracting $|V_{ub}|$ from inclusive semileptonic decays
$b \to u l \nu_l$ by applying a cut on the invariant mass of the
leptons $q^2$. To eliminate the charm background, one requires $q^2 >
q_0^2 = (m_B - m_D)^2 \approx 11.6~{\rm GeV}^2$. It turns out that
this cut is mild enough to keep significant fraction of $b \to u$
transitions and also the energy release is sufficiently large so that
the standard methods of the Heavy Quark Expansion in $1/m_b$ can be
applied with confidence.  Of course, there are several sources of the
theoretical uncertainties associated with this method, including in
roughly equal measure the value of the $b$ quark mass, the
non-perturbative power corrections (of third order in the ratio of
$\Lambda_{\rm QCD}$ and the characteristic momentum flow), and the
two-loop perturbative QCD corrections \cite{Neubert:2001ib}.  The
calculation of this last effect is the main purpose of this Letter.

The difficulty connected with such corrections is that they involve
the $q^2$ distribution, rather than the total decay rate.  While
two-loop corrections to charged particle decays are in general
challenging (the first calculations for specific kinematic
configurations or the total decay rates have been completed only
recently, see e.g.~\cite{zerorecoil,Czarnecki:1997hc,%
vanRitbergen:1998yd,vanRitbergen:1999gs}), two-loop corrections to the
decay distributions have never been evaluated so far.

\begin{figure}[htb]
\hspace*{-38mm}
\begin{minipage}{16.cm}
\begin{tabular}{cc}
\psfig{figure=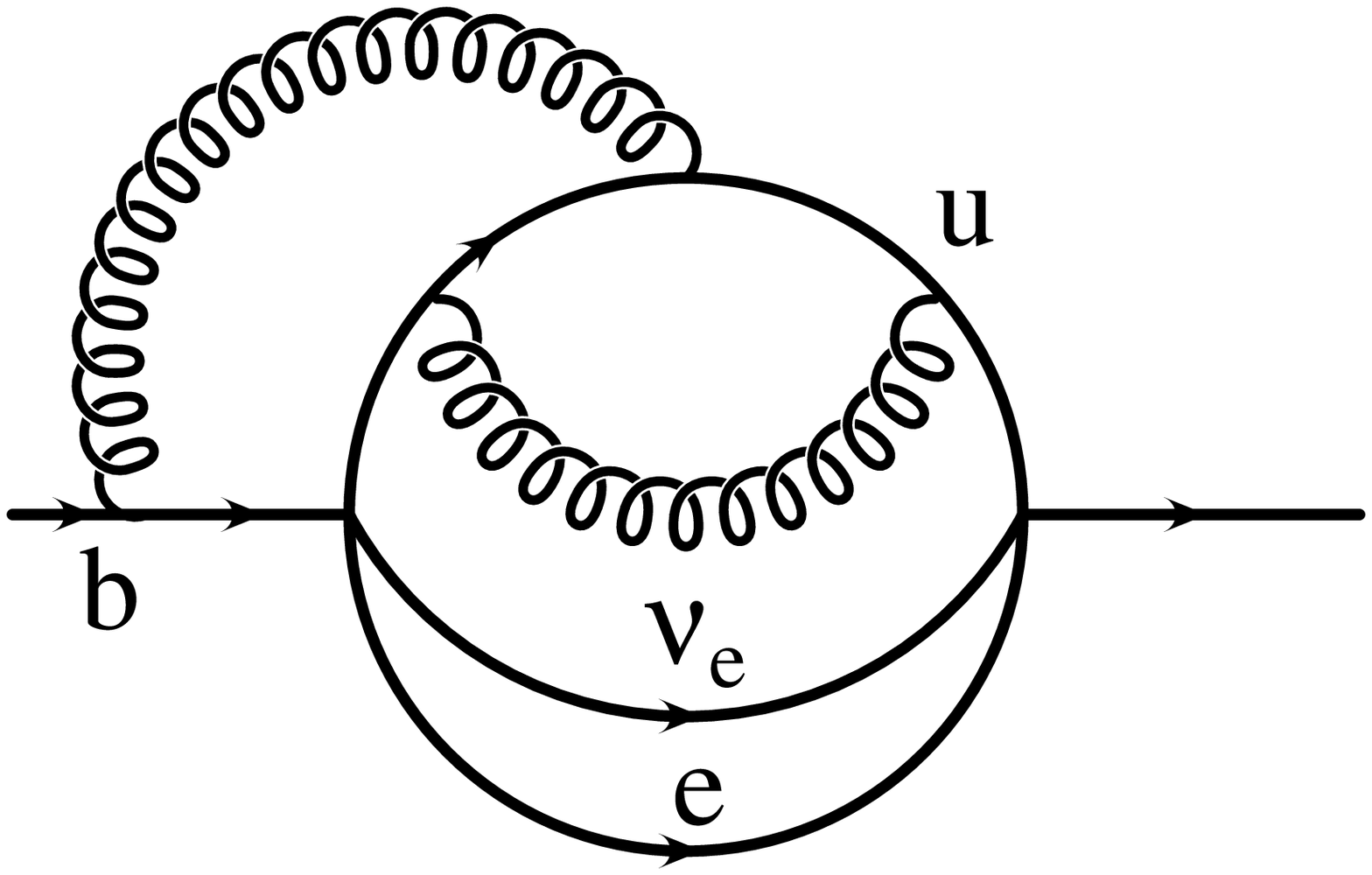,width=40mm}
&\hspace*{0mm}
\psfig{figure=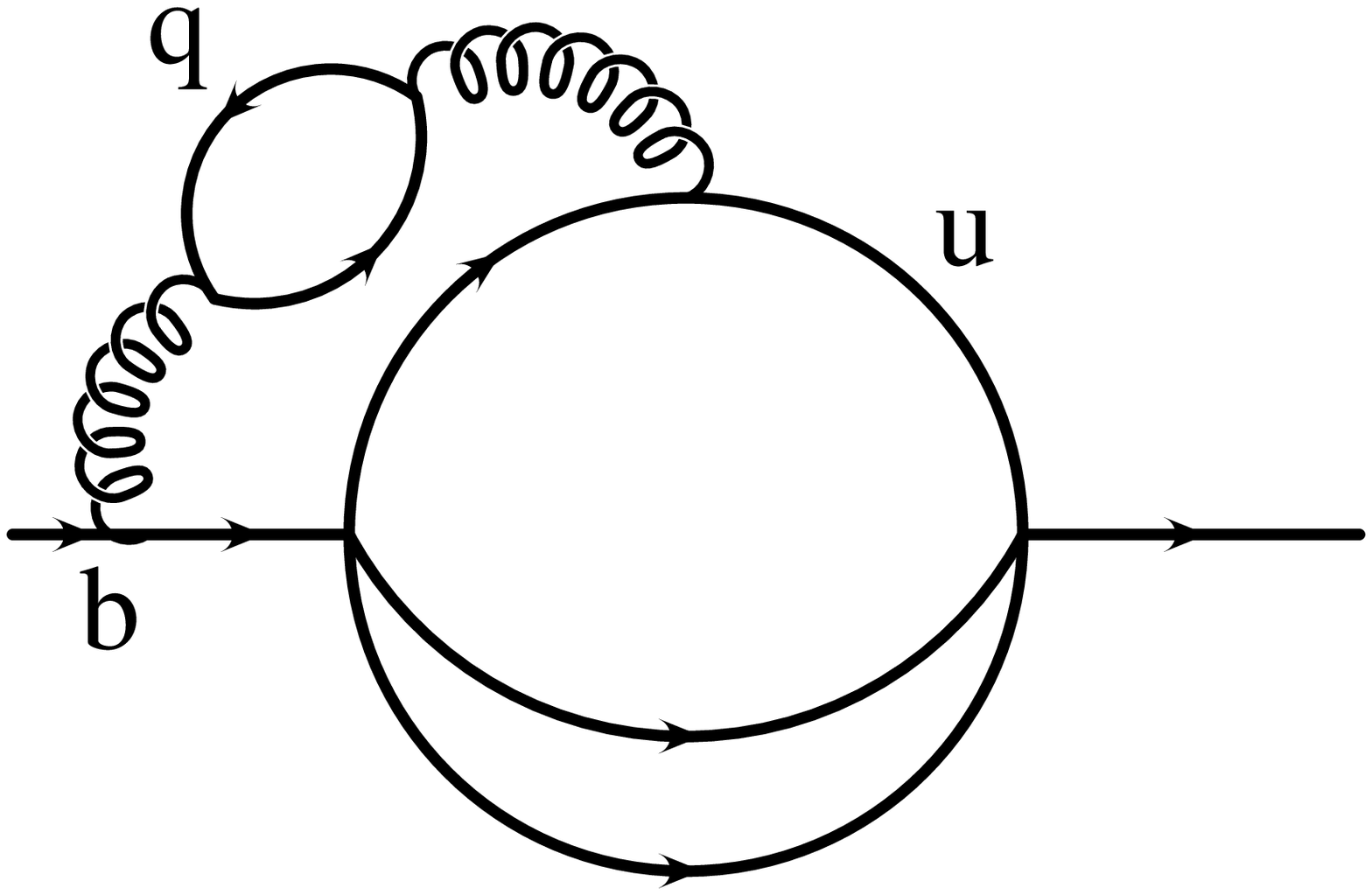,width=40mm}
\\[1mm]
(a) & (b)\\
\\[1mm]
\multicolumn{2}{c}{\psfig{figure=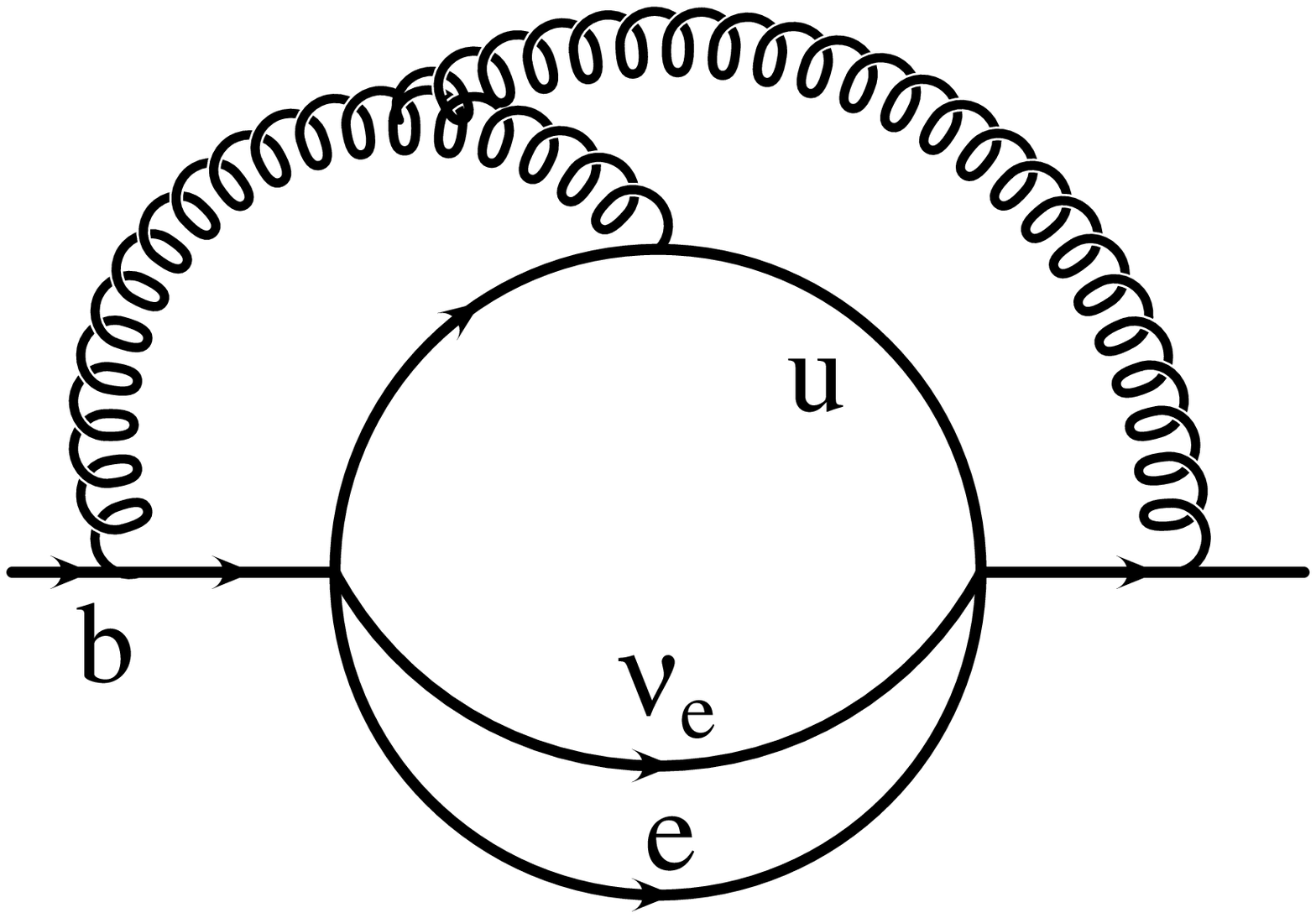,width=40mm}}
\\
\multicolumn{2}{c}{(c)}
\end{tabular}
\end{minipage}
\caption{Examples of diagrams whose cuts contribute to the
semileptonic decay $b\to u \,l\, \nu_l$: (a) abelian; (b) light or
heavy quarks; (c) non-abelian.}
\label{fig:loops}
\end{figure}

In the present calculation we take advantage of the fact that, for the
experimentally interesting case, the invariant mass of the leptons is
large.  We introduce an expansion parameter $\delta = (m_b^2 -
q^2)/m_b^2$.  In $b\to u$ studies using cuts proposed in
\cite{Bauer:2000xf,Bauer:2001rc} the maximal value of $\delta$ is
about 0.5 for $q^2 = q_0^2$.  Obviously, increasing $q^2$ results in a
rapid decrease of $\delta$, so that $\delta$ can be considered as a
small parameter in the region of interest $q^2 > q_0^2$.  Therefore,
by constructing an algorithm for expanding the relevant Feynman
diagrams around $\delta=0$ and computing several terms of such an
expansion, we can derive the ${\cal O}(\alpha_s^2)$ correction to the
dilepton invariant mass spectrum valid in the region of experimental
interest.

Examples of diagrams we have to consider in studying the semileptonic
$b\to u$ decay at $\order{\alpha_s^2}$ are shown in
Fig.~\ref{fig:loops}.  The optical theorem connects the imaginary part
of such diagrams with contributions to the decay.  We first integrate
over the lepton and neutrino phase space, thereby reducing the problem
to the decay $b \to W^*(q^2) u$, where $W^*$ is a virtual $W$ boson
with an invariant mass $q^2$.  In the limit $\delta \to 0$, $q^2$
approaches $m_b^2$.  Therefore, due to phase space constraints, $W^*$
becomes static.  The expansion in $\delta$ is constructed by applying
the Heavy Quark/Boson Expansion to the Feynman diagrams. The only
unusual feature in our case is that the initial $b$-quark is on the
mass shell. In the HQET limit, this leads to propagators of the type
$1/(2pk)$, whereas the $W^*$ boson is off-shell so that its propagator
has the form $1/(2pk + \delta)$.

Since we are interested in the $\order{\alpha_s^2}$ corrections to the
decay distributions, we have to consider the three-loop diagrams of
the self-energy type, like those shown in Fig.~\ref{fig:loops} and
extract their imaginary parts.  Initially, there are two scales in the
problem: using $m_b$ as a unit of energy, these scales can be
expressed as $\order{1}$ and $\order{\delta}$.  We employ asymptotic
expansions to indentify contributions arising from these widely
separated scales.  The region with all loop momenta of $\order{1}$
does not contribute to the imaginary part since it is analytic
(polynomial) in $\delta$.  When some loop momenta are $\order{
\delta}$ and others are $\order{1}$, a three-loop diagram factorizes
into a product of one- and/or two-loop diagrams and is easy to
evaluate.

The non-trivial part of the calculation is the HQET limit where all
loop momenta are of $\order{\delta}$. These diagrams are similar to
the three-loop HQET diagrams
\cite{Grozin:2000jv,Grozin:2001fw,Czarnecki:2001rh} but not identical
with them, since some of the lines in the present case are
on-shell. We have constructed an algorithm based on recurrence
relations and integration-by-parts identities \cite{che81} with which
one can reduce any relevant three-loop diagram to a linear combination
of a few master integrals. Four of these master integrals are new. We
compute them in the Euclidean ($p^2=-1$), $D=4-2\ep$ dimensional
space.

Propagators occuring in the master integrals are denoted by
$D_1 = k_1^2,~D_2=k_2^2,~D_3 = k_3^2,~ 
D_4 = (k_1-k_2)^2,~D_5 =(k_2-k_3)^2,~D_6=2pk_1,~D_7 = 2pk_2,
~D_8 = 2pk_3,
~D_9=2pk_1+2pk_2,~
D_{10} = 2pk_1+2pk_3,~D_{11} = 2pk_1 + 2pk_2+2pk_3$.  The four new
results are
\ba
\lefteqn{I_1 = \int \frac{{\rm d}^D k_1 \; {\rm d}^D k_2}{D_1^{\ep} D_2 
D_4 D_6 (D_7 + 1)} 
= C_\ep^2 \left [ -\frac{1}{24\ep^2} - \frac{5}{24\ep} \right.}
\nonumber \\
&& 
-\frac {13}{24} - \frac{17 \pi^2}{48}
+\ep \left ( \frac{23}{24}-\frac{85\pi^2}{48}
-\frac{11}{6}\zeta_3 \right )
\nonumber \\
&& \left.
+\ep^2 \left ( \frac{623}{24}-\frac{55}{6}\zeta_3-\frac{95}{16}\pi^2
 -\frac {16523 \pi^4}{8640} \right ) +\order{\ep^3} \right ],
\nonumber 
\ea
\vspace*{-5mm}
\ba
I_2 &=& \int \frac{{\rm d}^D k_1 \; {\rm d}^D k_2 \; {\rm d}^D
k_3}{D_1 D_2 D_3  
D_9 D_{10} (D_{11}+1) } 
 \nonumber \\
&=& C_\ep^3 \left [
        \frac {1}{\ep} \left (  - \frac {5}{18} + \frac {\pi^2}{36} \right )
 - \frac {9}{2} + \frac {\pi^2}{6} + \frac {7}{3}\zeta_3
\right.
 \nonumber \\
&&
\qquad
 \left.
 + \ep  \Big (  - \frac {91}{2} 
       - \frac {\pi^2}{6} + \frac {11 \pi^4}{36} + 14\zeta_3 \Big )
 +\order{\ep^2}
\right ],
\nonumber \\
I_3 &=& \int \frac{{\rm d}^D k_1 \; {\rm d}^D k_2 \; {\rm d}^D
k_3}{D_1 D_3 D_4 D_5 
D_6 D_7 (D_8+1) }
\nonumber \\
 &=& C_\ep^3 \left [
-\frac {\pi^2}{36 \ep^2} 
 + \frac {1}{\ep} \left ( \frac{2}{3}\zeta_3-\frac{2\pi^2}{9}\right ) 
\right. \nonumber \\
&& \left.
\qquad
+ \frac{16}{3}\zeta_3- \frac {13\pi^2}{9}-\frac{47\pi^4}{270}
 +\order{\ep}
\right ],
\nonumber \\
I_4 &=& \int \frac{{\rm d}^D k_1 \; {\rm d}^D k_2 \; {\rm d}^D
k_3}{D_1 D_3 D_4 D_5 
D_6 (D_7+1) D_8 } 
\nonumber \\
&=& C_\ep^3 \left [
-\frac {\zeta_3}{\ep} -8\zeta_3-\frac {\pi^4}{60} 
 +\order{\ep}
\right ].
\ea
In the above formulas $\zeta_3$ is the Riemann zeta function,
$\zeta_3 = \sum \limits_{i=1}^{\infty} 1/i^3$, and 
$C_\ep \equiv \pi^{2-\ep}\Gamma(1+\ep)$.

Using recurrence relations to reduce all loop integrals to a
combination of master integrals (these algebraic manipulations are
done with FORM \cite{form3}), we obtain the ${\cal O}(\alpha_s^2)$
correction to the dilepton invariant mass spectrum (we use the pole
mass $m_b$ and the $\overline{\rm MS}$ scheme for $\alpha_s$)
\be
\frac{1}{\Gamma_0} \frac {{\rm d}\Gamma}{{\rm d}|\delta|}
= 6\delta^2 - 4\delta^3 
+ \frac{\alpha_s(m_b)}{\pi}\, X_1 
+ \left (\frac{\alpha_s}{\pi} \right )^2  X_2,
\label{rate}
\ee
where $\Gamma_0 = G_F^2|V_{ub}|^2 m_b^5/192\pi^3$ and $X_{1,2}$ denote
the one- \cite{Jezabek:1993wk} 
and two-loop corrections, respectively, 
\ba
\lefteqn{\hspace*{-1.1mm}
X_1= C_F \left [
\delta^2 \left ( \frac{27}{2} - 9 L - 4\pi^2 \right )
+\delta^3 \left ( \frac{2}{3} - 2 L + \frac{8}{3}\pi^2 \right )
\right.}
 \nonumber \\
&& \left. +\delta^4 \left ( L - \frac {13}{3} \right)
- \frac {19}{30} \delta^5 
-\delta^6 \left(  \frac {31}{180} + \frac{L}{6} \right)
+\order{\delta^7}
\right ], \nonumber 
\\
\lefteqn{\hspace*{-1.1mm}
X_2 = C_F \left ( C_F X_{\rm A} + C_A X_{\rm NA} + T_R N_L X_{\rm L} 
+ T_R N_H X_{\rm H} \right )}
\label{defx}
\ea
where $L=\ln\delta$ and $C_F=4/3$, $C_A=3$, and $T_R=1/2$ are the
usual SU(3) color factors and $N_L$ and $N_H$ denote the number of light
($m_q=0$) and heavy ($m_q=m_b$) quark species.  We use the
approximation $m_c = m_b$ since for $q^2 > q_0^2$ there is no phase
space available for charm quark production. If needed, corrections for
$m_c \neq m_b$ in virtual effects can easily be computed.

For the coefficients $X_{\rm A},~X_{\rm NA},~X_{\rm L},~X_{\rm H}$ we
find
\ba
\lefteqn{ X_{\rm A} = \delta^2 \Bigg\{
      \frac{27}{4} L^2   - \left ( \frac{147}{8} -4\pi^2 \right )L 
     +\frac{523}{16}  - \frac{39}{2}\zeta_3}
\nonumber \\
&&\qquad + 7 \pi^2 \ln 2 
     - \frac{71 \pi^2}{6} + \frac{16 \pi^4}{15}
\nonumber \\
&&  
 + \delta \left[ \frac {15}{2} L^2 - \left (\frac{287}{12} 
      - \frac{10 \pi^2}{9} \right ) L
\right. \nonumber \\
&& \left. \qquad 
+ \frac {1363}{72} - \frac{2}{3}\pi^2 \ln 2 - \frac{217}{108}\pi^2 
       - \frac{32 \pi^4}{45} \right]
\nonumber \\
&& + \delta^2 \left[ \frac  {17}{4} L^2
          - \left (\frac{67}{6} + \frac{17 \pi^2}{18} \right ) L 
                       + \frac{1537}{144} 
\right.
 \nonumber \\
&& \left.
\qquad   - \frac {5}{6}\pi^2\ln 2  + \frac {937 \pi^2}{432} \right ]
 \nonumber \\
&&
 + \delta^3 \left[ \frac{11}{3}L^2 - \left ( 
         \frac{269}{45} + \frac{8 \pi^2}{45} \right ) L
          + \frac{43609}{21600}
\right. \nonumber \\
&& \left.
\qquad
 - \frac{1}{2}\zeta_3 - \frac{\pi^2}{5}\ln 2 
 + \frac{7609 \pi^2}{10800} \right]
 \nonumber \\
&&
       + \delta^4 \left[ \frac{77}{30}L^2 - \frac{3071}{600}L 
   +\frac{579137}{162000} - \frac{17}{40}\zeta_3
\right. \nonumber \\
&& \left.
\qquad
  - \frac{\pi^2}{20}\ln 2 
 - \frac{451}{1080}\pi^2 \right] \Bigg\},
\nonumber \\
\lefteqn{ X_{\rm NA} = \delta^2 \Bigg\{
\frac{33}{4} L^2  - \left ( \frac{423}{8} - \frac{47\pi^2}{6} \right ) L
         + \frac{1103}{16}  }
\nonumber \\
&&
\qquad
 - \frac {129}{4}\zeta_3
- \frac{7\pi^2}{2}\ln 2 - \frac{881 \pi^2}{72} + \frac{13 \pi^4}{30} 
\nonumber \\
&&
 + \delta \left[\frac {4}{3} L^2 
      + \left ( \frac {155}{36} - 4 \pi^2 \right ) L
\right. \nonumber \\
&& \left.  
\qquad
- \frac{623}{27}  + 24\zeta_3 
  + \frac{\pi^2}{3} \ln 2 + \frac{247 \pi^2}{24} - \frac{13 \pi^4}{45}
\right] 
\nonumber \\
&& 
+ \delta^2 \left[   -\frac{17}{12} L^2 
     + \left (\frac{1877}{144} - \frac{\pi^2}{12} \right ) L
   - \frac{18319}{864} 
\right. \nonumber \\
&& \left. 
\qquad
          - \frac{1}{4}\zeta_3 
+ \frac{5 \pi^2}{12}\ln 2 
    - \frac{19 \pi^2}{288} \right]
\nonumber \\
&&
 + \delta^3 \left[ -\frac{11}{24} L^2 
        + \left (\frac{139}{45} + \frac {\pi^2}{45} \right ) L
\right. \nonumber \\
&& \left.
\qquad - \frac {52379}{21600} 
         -\frac {\zeta_3}{20} + \frac{\pi^2}{10}\ln 2 
     - \frac{5041 \pi^2}{21600} \right]
\nonumber \\
&&
+ \delta^4 \left[  \frac{2}{15} L^2 
+ \left (\frac{36433}{21600} 
 + \frac{7 \pi^2}{180}  \right ) L
- \frac{1025099}{648000}  
\right. \nonumber \\
&& \left.
\qquad 
        + \frac{1}{80}\zeta_3 
+ \frac{\pi^2}{40}\ln 2 
 + \frac{1243 \pi^2}{5400} \right]
\Bigg\},
\nonumber 
\\
\lefteqn{X_{\rm L} = \delta^2 \Bigg\{
 - 3 L^2 + \left ( \frac{39}{2} - \frac{8 \pi^2}{3} \right ) L
  - \frac {117}{4}}
\nonumber \\
&&
\qquad  + 12\zeta_3 + \frac{41 \pi^2}{9}  
\nonumber \\          
&& + \delta \left[ - \frac{2}{3} L^2
  -\left (\frac{31}{9} -\frac{16\pi^2}{9} \right )L
+ \frac{797}{54} \right.
\nonumber \\ &&
\left.
 \qquad  - 8\zeta_3 - \frac{106\pi^2}{27} \right ]
\nonumber \\  
&&       + \delta^2 \left[ \frac{1}{3} L^2 -\frac{37}{9}L
 + \frac{1289}{216} + \frac{\pi^2}{9} \right]
\nonumber \\  
&& 
       + \delta^3 \left[ - \frac{6}{5} L + \frac{1817}{1800}  \right]
\nonumber \\  
&& 
       + \delta^4 \left[ -\frac{1}{18}L^2 - \frac{98}{135} L 
+ \frac{8129}{16200}  - \frac{\pi^2}{54} \right]
\Bigg\}.
\nonumber \\
X_{\rm H} &=&\delta^2\Bigg[ {133\over 8} - {5\over 3} \pi^2 
+ \delta \left(  - {797\over 108 } + { 2\over 3} \pi^2 \right)
\nonumber \\
&&+ \delta   \left( {2473\over 2700} - {1\over 18 } \pi^2 - {1\over 5}
L\right) 
\nonumber \\
&&
+ \delta^2 \left( {1747\over 2700} - {1\over 15 } \pi^2 - {2\over 15}
L\right) 
\nonumber \\
&&
+ \delta^3 \left( {816239\over 1587600} - {29\over 540 } \pi^2
              - {187\over 1260} L 
\right)
\nonumber \\
&&+ \delta^4 \left( {36059\over 88200} - {3\over 70 } \pi^2
  - {103\over 630} L \right)\Bigg].
\ea

For brevity we have presented the results accurate up to the terms
${\cal O}(\delta^6)$.  For the numerical analysis below we use terms
up to ${\cal O}(\delta^8)$.

We tested these results in several ways.  We used a general covariant
gauge and checked the cancellation of the gauge parameter.  The result
for $X_{\rm L}$ agrees with the numerical calculation in
\cite{Luke:1995du}. A simple interpolating formula which we actually
used for the comparison can be found in the appendix of
Ref.~\cite{Bauer:2001rc}.  The agreement is very good, practically for
all values of $\delta$.

Further, we can extrapolate the results of the expansion by taking the
limit $\delta \to 1$ in which case our formulas should describe the
decay of a massive quark into a massless quark and a massless $W$
boson. In this limit, second order QCD corrections were computed for
the top quark decay \cite{Czarnecki:1998qc}.  We find that for the
color structures $X_{\rm NA,L,H}$ the difference between the two
results is better than $10\%$.  The agreement is much worse for the
abelian part $X_{\rm A}$, where the difference can be as large as
$50\%$.  This demonstrates that the seven terms of the expansion are
insufficient for the abelian part to converge in the limit $\delta \to
1$.

  However, because of the SU(3) color factors, the contribution of the
abelian part is suppressed and we can reliably derive the ${\cal
O}(\alpha_s^2)$ correction to top quark decay from our formulas.
Taking $N_L=5$ and $N_H=1$, we find $X_2/2 \approx -16.4$, whereas the
central values of the coefficients in \cite{Czarnecki:1998qc} give
$-16.7$.  An even better agreement is obtained for $\delta=
\delta_W\equiv 1-M_W^2/m_t^2 \simeq 0.79$.  At this point,
corresponding to physical values of the $W$ boson and top quark
masses, the width of $t\to bW$ was evaluated in
\cite{Chetyrkin:1999ju}.  We have perfect agreement with the central
value of the second order correction, $X_2(\delta_W)/2 = -15.6$, given
in eq.~(28) of that paper.

\begin{figure}[htb]
\hspace*{2 mm}
\begin{minipage}{16.cm}
\begin{picture}(100,0)
\put (-185,-10) {$X_2$}
\put (50,-135) {$\delta$}
\end{picture}

\psfig{figure=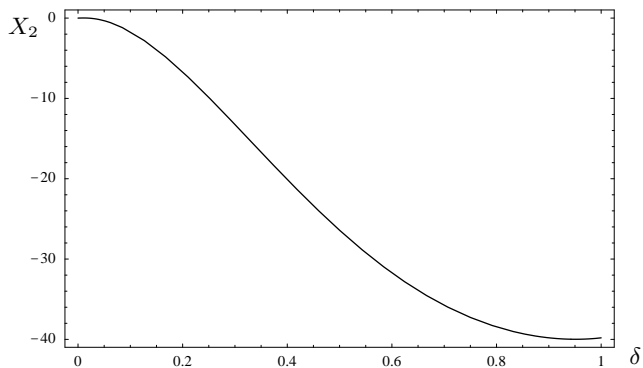,width=78mm}
\end{minipage}
\caption{ $\order{\alpha_s^2}$ correction to the decay width $b\to
u l \nu_l$, $X_2(\delta)$ (defined in eq.~(\protect\ref{defx})), 
as a function of $\delta = 1-q^2/m_b^2$ (for $N_L=3$, $N_H=2$).}
\label{fig:plot}
\end{figure}

As the final check one can integrate Eq.~(\ref{rate}) over $\delta$,
obtaining the total decay rate $b \to u l \nu_l$, for which the second
order QCD corrections are known \cite{vanRitbergen:1999gs}.  Taking
$N_L=4$ and $N_H=1$ and integrating over $\delta$ we obtain
$\int_{0}^{1} {\rm d} \delta\, X_2(\delta) = -21.24$, in excellent
agreement with $-21.296$, given in Ref. \cite{vanRitbergen:1999gs}.

Integrating the abelian contribution $X_{\rm A}$ we can compute the
two-photon corrections to the muon lifetime.  We find $\int_{0}^{1}
{\rm d} \delta\, X_{\rm A}(\delta) \simeq 3.1$, where the 13\%
discrepancy with the exact value in eq.~(9) of
\cite{vanRitbergen:1998yd} is due to poor convergence of our series
for large $\delta$.  However, if we assume that the convergence is
good up to $\delta\approx 0.65$ and extrapolate for larger $\delta$
using $X_{\rm A}(1)=7.0(4)$ \cite{Czarnecki:1998qc}, we reproduce the
muon lifetime correction \cite{vanRitbergen:1998yd} within 3\%.

For $\delta \lsim 1/2$, relevant for the extraction of $|V_{ub}|$, the
series converge very well and accurately approximate all color
components of the $\order{\alpha_s^2}$ correction.

The full $\order{\alpha_s^2}$ correction to the quark decay width,
$X_2(\delta)$, is plotted in Fig.~\ref{fig:plot}.  Even at the end
point $\delta=1$, our estimate for $X_2$ agrees with our result for
the top decay \cite{Czarnecki:1998qc} to better than $3\%$.

To show the impact of the computed corrections on  dilepton invariant mass 
distribution, we  separate the BLM \cite{BLM} 
and non-BLM corrections since the 
former have already been studied in the literature. We define the BLM
and non-BLM
corrections as
\be
X_2^{\rm BLM} = -3C_F \beta_0 X_L, \qquad X_2^{\rm nonBLM} = X_2 -
X_2^{\rm BLM},
\ee
where $\beta_0 = 11C_A/12 - T_R N_L/3$ denotes the  beta-function
coefficient 
in a theory with three massless quark flavors, appropriate 
for the range  of $q^2$  used for the $V_{ub}$ extraction.

\begin{figure}[htb]
\vspace*{8 mm}
\hspace*{2 mm}
\begin{minipage}{16.cm}
\begin{picture}(100,0)
\put (-185,2) {$X_2^{\rm nonBLM}$}
\put (50,-139) {$\delta$}
\end{picture}

\psfig{figure=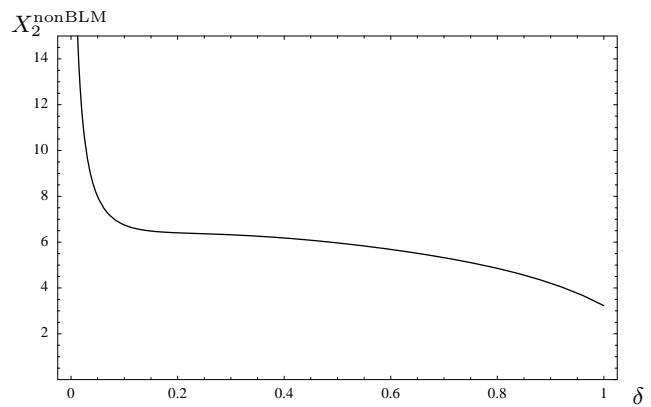,width=78mm}
\end{minipage}
\caption{The non-BLM corrections, $X_2^{\rm
nonBLM}/(6\delta^2-4\delta^3)$ (for $N_L=3$, $N_H=2$).} 
\label{nonBLM}
\end{figure}

The value of the BLM corrections is known to be strongly correlated
with the scale of the coupling constant used in the one-loop result
and also with the quark mass used in the formula for the decay rate A
discussion of these issues can be found in the literature
\cite{Bauer:2001rc} and we will not consider them here.  On the
contrary, the non-BLM corrections are new.  Their dependence on
$\delta$ is shown in Fig.~\ref{nonBLM} where the ratio of the non-BLM
corrections and the tree level decay rate $6 \delta^2 - 4 \delta^3$ is
plotted. For realistic values of the strong coupling constant,
$\alpha_s = 0.2 - 0.3$, the non-BLM corrections are about $5\%$ in the
range of $\delta$ relevant for the $|V_{ub}|$ extraction from the
dilepton invariant mass spectrum.

The technique described in this Letter might open a way to reliable
estimates of the $\order{\alpha_s^2}$ corrections to more complicated
observables.  For example, a simple modification allows one to
calculate the moments of the charged lepton energy spectrum for a
fixed value of the dilepton invariant mass.

Recently, combined cuts on both dilepton and hadron invariant masses
were advocated for the $|V_{ub}|$ determination \cite{Bauer:2001rc}.
It has been argued that in this approach one can keep the
theoretical uncertainties under control while retaining a larger data
sample of the $b \to u$ transitions.  Since the calculation reported
here has been performed without any restriction on the hadronic
invariant mass, our results for the QCD corrections are not applicable
in this case.  However, a sufficiently large number of moments should
contain enough information about the spectrum to determine the effect
of the cut.

{\em Acknowledgments:} 
This research was supported in part by the Natural
Sciences and Engineering Research Council of Canada and by the DOE
under grant number DE-AC03-76SF00515.


\begin{thebibliography}{17}
\expandafter\ifx\csname natexlab\endcsname\relax\def\natexlab#1{#1}\fi
\expandafter\ifx\csname bibnamefont\endcsname\relax
  \def\bibnamefont#1{#1}\fi
\expandafter\ifx\csname bibfnamefont\endcsname\relax
  \def\bibfnamefont#1{#1}\fi
\expandafter\ifx\csname citenamefont\endcsname\relax
  \def\citenamefont#1{#1}\fi
\expandafter\ifx\csname url\endcsname\relax
  \def\url#1{\texttt{#1}}\fi
\expandafter\ifx\csname urlprefix\endcsname\relax\def\urlprefix{URL }\fi
\providecommand{\bibinfo}[2]{#2}
\providecommand{\eprint}[2][]{\url{#2}}

\bibitem[{\citenamefont{Bauer et~al.}(2000)\citenamefont{Bauer, Ligeti, and
  Luke}}]{Bauer:2000xf}
\bibinfo{author}{\bibfnamefont{C.~W.} \bibnamefont{Bauer}},
  \bibinfo{author}{\bibfnamefont{Z.}~\bibnamefont{Ligeti}}, \bibnamefont{and}
  \bibinfo{author}{\bibfnamefont{M.~E.} \bibnamefont{Luke}},
  \bibinfo{journal}{Phys. Lett.} \textbf{\bibinfo{volume}{B479}},
  \bibinfo{pages}{395} (\bibinfo{year}{2000}),
  \eprint[http://arXiv.org/abs]{hep-ph/0002161}.

\bibitem[{\citenamefont{Bauer et~al.}(2001)\citenamefont{Bauer, Ligeti, and
  Luke}}]{Bauer:2001rc}
\bibinfo{author}{\bibfnamefont{C.~W.} \bibnamefont{Bauer}},
  \bibinfo{author}{\bibfnamefont{Z.}~\bibnamefont{Ligeti}}, \bibnamefont{and}
  \bibinfo{author}{\bibfnamefont{M.~E.} \bibnamefont{Luke}},
  \bibinfo{journal}{Phys. Rev.} \textbf{\bibinfo{volume}{D64}},
  \bibinfo{pages}{113004} (\bibinfo{year}{2001}),
  \eprint[http://arXiv.org/abs]{hep-ph/0107074}.

\bibitem[{\citenamefont{Neubert and Becher}(2001)}]{Neubert:2001ib}
\bibinfo{author}{\bibfnamefont{M.}~\bibnamefont{Neubert}} \bibnamefont{and}
  \bibinfo{author}{\bibfnamefont{T.}~\bibnamefont{Becher}}
  (\bibinfo{year}{2001}), \eprint[http://arXiv.org/abs]{hep-ph/0105217}.

\bibitem[{\citenamefont{Czarnecki}(1996)}]{zerorecoil}
\bibinfo{author}{\bibfnamefont{A.}~\bibnamefont{Czarnecki}},
  \bibinfo{journal}{Phys. Rev. Lett.} \textbf{\bibinfo{volume}{76}},
  \bibinfo{pages}{4124} (\bibinfo{year}{1996}).

\bibitem[{\citenamefont{Czarnecki and Melnikov}(1997)}]{Czarnecki:1997hc}
\bibinfo{author}{\bibfnamefont{A.}~\bibnamefont{Czarnecki}} \bibnamefont{and}
  \bibinfo{author}{\bibfnamefont{K.}~\bibnamefont{Melnikov}},
  \bibinfo{journal}{Phys. Rev. Lett.} \textbf{\bibinfo{volume}{78}},
  \bibinfo{pages}{3630} (\bibinfo{year}{1997}), \eprint{hep-ph/9703291}.

\bibitem[{\citenamefont{van Ritbergen and Stuart}(1999)}]{vanRitbergen:1998yd}
\bibinfo{author}{\bibfnamefont{T.}~\bibnamefont{van Ritbergen}}
  \bibnamefont{and} \bibinfo{author}{\bibfnamefont{R.~G.}
  \bibnamefont{Stuart}}, \bibinfo{journal}{Phys. Rev. Lett.}
  \textbf{\bibinfo{volume}{82}}, \bibinfo{pages}{488} (\bibinfo{year}{1999}),
  \eprint{hep-ph/9808283}.

\bibitem[{\citenamefont{van Ritbergen}(1999)}]{vanRitbergen:1999gs}
\bibinfo{author}{\bibfnamefont{T.}~\bibnamefont{van Ritbergen}},
  \bibinfo{journal}{Phys. Lett.} \textbf{\bibinfo{volume}{B454}},
  \bibinfo{pages}{353} (\bibinfo{year}{1999}),
  \eprint[http://arXiv.org/abs]{hep-ph/9903226}.

\bibitem[{\citenamefont{Grozin}(2000)}]{Grozin:2000jv}
\bibinfo{author}{\bibfnamefont{A.~G.} \bibnamefont{Grozin}},
  \bibinfo{journal}{JHEP} \textbf{\bibinfo{volume}{03}}, \bibinfo{pages}{013}
  (\bibinfo{year}{2000}), \eprint{hep-ph/0002266}.

\bibitem[{\citenamefont{Grozin}(2001)}]{Grozin:2001fw}
\bibinfo{author}{\bibfnamefont{A.~G.} \bibnamefont{Grozin}}
  (\bibinfo{year}{2001}), \eprint{hep-ph/0107248}.

\bibitem[{\citenamefont{Czarnecki and Melnikov}(2001)}]{Czarnecki:2001rh}
\bibinfo{author}{\bibfnamefont{A.}~\bibnamefont{Czarnecki}} \bibnamefont{and}
  \bibinfo{author}{\bibfnamefont{K.}~\bibnamefont{Melnikov}}
  (\bibinfo{year}{2001}), \eprint[http://arXiv.org/abs]{hep-ph/0110028}.

\bibitem[{\citenamefont{Chetyrkin and Tkachev}(1981)}]{che81}
\bibinfo{author}{\bibfnamefont{K.~G.} \bibnamefont{Chetyrkin}}
  \bibnamefont{and} \bibinfo{author}{\bibfnamefont{F.~V.}
  \bibnamefont{Tkachev}}, \bibinfo{journal}{Nucl. Phys.}
  \textbf{\bibinfo{volume}{B192}}, \bibinfo{pages}{159} (\bibinfo{year}{1981}).

\bibitem[{\citenamefont{Vermaseren}()}]{form3}
\bibinfo{author}{\bibfnamefont{J.~A.~M.} \bibnamefont{Vermaseren}},
  \bibinfo{note}{math-ph/0010025}.

\bibitem[{\citenamefont{Je{\.z}abek and K{\"u}hn}(1993)}]{Jezabek:1993wk}
\bibinfo{author}{\bibfnamefont{M.}~\bibnamefont{Je{\.z}abek}} \bibnamefont{and}
  \bibinfo{author}{\bibfnamefont{J.~H.} \bibnamefont{K{\"u}hn}},
  \bibinfo{journal}{Phys. Rev.} \textbf{\bibinfo{volume}{D48}},
  \bibinfo{pages}{1910} (\bibinfo{year}{1993}), \bibinfo{note}{e: ibid. {\bf
  D49}, 4970 (1994)}, \eprint{hep-ph/9302295}.

\bibitem[{\citenamefont{Luke et~al.}(1995)\citenamefont{Luke, Savage, and
  Wise}}]{Luke:1995du}
\bibinfo{author}{\bibfnamefont{M.~E.} \bibnamefont{Luke}},
  \bibinfo{author}{\bibfnamefont{M.~J.} \bibnamefont{Savage}},
  \bibnamefont{and} \bibinfo{author}{\bibfnamefont{M.~B.} \bibnamefont{Wise}},
  \bibinfo{journal}{Phys. Lett.} \textbf{\bibinfo{volume}{B343}},
  \bibinfo{pages}{329} (\bibinfo{year}{1995}),
  \eprint[http://arXiv.org/abs]{hep-ph/9409287}.

\bibitem[{\citenamefont{Czarnecki and Melnikov}(1999)}]{Czarnecki:1998qc}
\bibinfo{author}{\bibfnamefont{A.}~\bibnamefont{Czarnecki}} \bibnamefont{and}
  \bibinfo{author}{\bibfnamefont{K.}~\bibnamefont{Melnikov}},
  \bibinfo{journal}{Nucl. Phys.} \textbf{\bibinfo{volume}{B544}},
  \bibinfo{pages}{520} (\bibinfo{year}{1999}),
  \eprint[http://arXiv.org/abs]{hep-ph/9806244}.

\bibitem[{\citenamefont{Chetyrkin et~al.}(1999)\citenamefont{Chetyrkin,
  Harlander, Seidensticker, and Steinhauser}}]{Chetyrkin:1999ju}
\bibinfo{author}{\bibfnamefont{K.~G.} \bibnamefont{Chetyrkin}},
  \bibinfo{author}{\bibfnamefont{R.}~\bibnamefont{Harlander}},
  \bibinfo{author}{\bibfnamefont{T.}~\bibnamefont{Seidensticker}},
  \bibnamefont{and}
  \bibinfo{author}{\bibfnamefont{M.}~\bibnamefont{Steinhauser}},
  \bibinfo{journal}{Phys. Rev.} \textbf{\bibinfo{volume}{D60}},
  \bibinfo{pages}{114015} (\bibinfo{year}{1999}),
  \eprint[http://arXiv.org/abs]{hep-ph/9906273}.

\bibitem[{\citenamefont{Brodsky et~al.}(1983)\citenamefont{Brodsky, Lepage, and
  Mackenzie}}]{BLM}
\bibinfo{author}{\bibfnamefont{S.~J.} \bibnamefont{Brodsky}},
  \bibinfo{author}{\bibfnamefont{G.~P.} \bibnamefont{Lepage}},
  \bibnamefont{and} \bibinfo{author}{\bibfnamefont{P.~B.}
  \bibnamefont{Mackenzie}}, \bibinfo{journal}{Phys. Rev.}
  \textbf{\bibinfo{volume}{D28}}, \bibinfo{pages}{228} (\bibinfo{year}{1983}).

\end{thebibliography}

\end{document}